\documentclass[3p]{elsarticle} 

\usepackage{setspace}

\usepackage{lineno,hyperref}

\usepackage[utf8]{inputenc} 
\usepackage{graphicx,psfrag,color}
\usepackage{subfig} 
\usepackage{caption} 
\usepackage{tabularx}
\usepackage{booktabs} 
\usepackage{multirow}
\usepackage{float}

\usepackage{amsmath} 
\usepackage{amssymb} 
\usepackage{amsfonts}
\usepackage{amsxtra}
\usepackage{tensor}
\usepackage{empheq}
\usepackage{stmaryrd}
\usepackage{icomma}
\usepackage{relsize}
\usepackage{trfsigns}
\usepackage[locale=US, unit-mode=text]{siunitx}

\usepackage{hyperref}

\newcommand{\M}[1]{\underline{\underline{\boldsymbol #1}}} 
\newcommand{\V}[1]{\underline{\boldsymbol #1}} 

\newcommand{\imag}{\text{j}}

\usepackage{tikz}
\usepackage{pgfplots}

\usepackage{array,booktabs,ragged2e}
\newcolumntype{R}[1]{>{\RaggedLeft\arraybackslash}p{#1}}
\newcolumntype{P}[1]{>{\centering\arraybackslash}p{#1}}

\usetikzlibrary{positioning}
\usetikzlibrary{backgrounds}
\tikzstyle{background rectangle}=[draw]

\DeclareUnicodeCharacter{2212}{--} 

\usepackage{array} 

\usepackage{xcolor} 

\begin{document}

\begin{frontmatter}

\title{Assessment of hybrid machine learning models for non-linear system identification of fatigue test rigs}

\author[IFKM]{L. Heindel}
\author[IFKM,DCFR]{P. Hantschke}
\author[IFKM,DCFR]{M. Kästner\corref{cor1}}

\address[IFKM]{Technische Universität Dresden, Institute of Solid Mechanics, 01062 Dresden, Germany}
\address[DCFR]{Dresden Center for Fatigue and Reliability (DCFR), 01062 Dresden, Germany}

\cortext[cor1]{Corresponding author. E-mail: Markus.Kaestner@tu-dresden.de ; Tel.: +49 351 463-43065 ; fax: +49 351 463-37061.}

\begin{abstract}
{\it

The prediction of system responses for a given fatigue test bench drive signal is a challenging task, for which linear frequency response function models are commonly used. To account for non-linear phenomena, a novel hybrid model is suggested, which augments existing approaches using Long Short-Term Memory networks. Additional virtual sensing applications of this method are demonstrated. The approach is tested using non-linear experimental data from a servo-hydraulic test rig and this dataset is made publicly available. A variety of metrics in time and frequency domains, as well as fatigue strength under variable amplitudes, are employed in the evaluation.}

\end{abstract}

\begin{keyword}
hybrid modeling \sep LSTM network \sep frequency response function model \sep forward prediction \sep virtual sensing
\end{keyword}

\end{frontmatter}


\section{Introduction}
In fatigue testing of vehicle components, a common problem is the accurate reproduction of service loads from measurement campaigns on a fatigue test bench. Since a control error always remains in practical, highly dynamic applications, it is necessary to adapt the drive signal, the input of the controller, in such a way that it leads to the desired system response after control has taken place. Due to the complexity of service loads, this process can not be conducted manually. In practice, a dynamic response simulation \cite{Dodds2001} is conducted, which involves system identification by means of a linear model of the dynamic system. Frequency response function (FRF) models have been used for test rig descriptions since 1976 \cite{Cryer1976}. Including further developments \cite{Hay2007}, they are state of the art for system identification and dynamic response simulation.\\
In order to account for the non-linear dynamics of fatigue test rigs, the error resulting from the linear FRF models is successively reduced in experimental iterations. These experiments can introduce significant pre-damage to the test specimen and lead to high energy consumption caused by the servo hydraulic test rigs. These problems could be greatly reduced by a non-linear forward prediction model, which approximates the system response for a given drive signal. With sufficient model accuracy, a suitable drive signal could be identified in simulations of the test bench, without necessitating experimental efforts for each trial drive signal. An experimental validation would only be required once a suitable drive signal is identified through simulation. As a result, the volume of experimental efforts could be significantly reduced.\\
However, the system identification task of forward prediction for fatigue test benches poses a set of specific requirements, which complicate the deployment of many classical approaches. These requirements are:
\begin{description}
	\item[Economic parameterization:] The number of available test specimens for a unique component test setup is generally rather low, therefore the process of system identification must be economical.
	\item[No intermediate results:] The entire drive signal must be defined before starting the experimental run. Therefore, intermediate measurement results of the predicted response quantity are not available for the simulation.
	\item[Fast and robust simulation:] In order to obtain a potential drive signal for the test bench, many potential input signals will need to be simulated. This process becomes impossible if the simulation itself is too slow or relies on an iterative scheme that can fail to converge.
	\item[Noise parameterization:] It is highly beneficial if the method can be parameterized using noise measurements in order to minimize the pre-damage to the specimen during parameterization.
\end{description}
As a result of these requirements, most commonly deployed strategies for non-linear system identification are not suited to forward prediction simulations of fatigue test rigs. A general review of classical methods is given in \textsc{Ljung} \cite{Ljung1999} and \textsc{Kerschen et al.} \cite{Kerschen2006}. Simulation approaches based on physical models generally require a very high effort for model building and parameterization when interactions between test bench and specimen, as well as non-linear and highly dynamic behavior are of interest. As a result, physics based models are not economic regarding the low number of fatigue test specimen. Kalman Filters \cite{Ljung1999, EftekharAzam2015} are well studied in the system identification literature and extensions for non-linear systems \cite{Lee1994} exist. However, since Kalman Filters inherently depend on a physical model of the system of interest, they are also not suited for forward prediction. There are also many variations of autoregressive models \cite{Ljung1999, Rouss2009} e.g. ARMA, ARMAS and NARX, which predict the system response of interest from its recent history as well as other available observed system inputs and outputs. Since no intermediate results of the predicted quantity are available during the test rig simulation, these models can also not be deployed. To the authors' knowledge, no non-linear forward prediction models are currently being deployed in dynamic response simulations, which fulfill these requirements.\\
While non-linear system identification is very challenging, it has more applications in fatigue analysis than just forward prediction. Virtual sensing aims to approximate unmeasured physical quantities in a system using existing concurrent sensor information. This can be highly beneficial when some quantities of interest are difficult to measure directly, but can be inferred from more accessible information. Example applications include structural health monitoring in general \cite{Hjelm2005,Ching2007,Erazo2014,Kullaa2019} as well as for off-shore structures \cite{Iliopoulos2014,Tarpoe2020}. Virtual sensing models also share two essential requirements with forward prediction tasks. Since no physical measurement of the quantity of interest is available, the history of that quantity can not be used for prediction. Further, simulations need to be fast and robust in order to efficiently process large amounts of measurement data in monitoring situations. The main difference compared to forward prediction is that more involved, physics based models are also suited for virtual sensing since the simulation models are deployed over significantly longer usage periods, warranting the increased parameterization efforts.\\
Recent breakthroughs of machine learning algorithms in image, audio and video processing brought forward novel data driven algorithms with great potential for application in forward prediction and virtual sensing. The SINDy model \cite{Brunton2016} employs sparse regression to identify the governing differential equations of a dynamic system from data. Physics constrained \cite{Kalina2023a,Linden2023} and physics informed neural networks (PINN) \cite{Raissi2019,SahliCostabal2020} have shown that existing knowledge about physical principles can be considered by machine learning models in order to achieve more realistic predictions. Finally, Long Short-Term Memory (LSTM) networks \cite{Hochreiter1997,Gers1999} have been successfully applied to a wide range of time series regression and classification tasks. They have proven to be powerful black box estimators, which can capture highly non-linear dynamic behavior. Of these three model classes, LSTM networks are the most promising candidates for the application to forward prediction tasks, since they fulfill all requirements listed above. While SINDy models are very well suited for system understanding, they rely on differential equation solvers to simulate system responses, which are slow and can fail to converge for noisy measurement data. PINN models are not suited for fatigue test applications, because the governing field equations involved in the dynamic system composed of test bench, controller and specimen are too complex for the incorporation of basic physical principals in the model prediction.\\
This paper shows how FRF models and LSTM networks can be combined to create novel hybrid models, which are well suited to forward prediction tasks. The LSTM network is used to modify the linear FRF predictions in order to account for system non-linearities, which is the most significant limitation of FRF models up to this point. The paper also shows that these hybrid models can be applied to virtual sensing tasks as well, which can be highly beneficial in situations where a physics-based model is not available. The content is structured as follows: In \autoref{sec:model_setup}, important fundamentals regarding frequency response function models and LSTM networks will be reviewed. Signal approximation as a combination of short subsequences is described and different hybrid modeling approaches are proposed. In \autoref{sec:experiments}, a variety of error metrics are introduced and the approach is tested on experimental data originating from a fatigue test bench. The paper concludes with a discussion of the results and an outlook in \autoref{sec:conclusions}.

\section{Material and methods} \label{sec:model_setup}
In this paper, vectors and matrices are indicated using single and double underscores respectively, e.g., $\V v$ and $\M M$. Whenever a scalar operator is applied to a vector, it represents an element-wise application of that operator. In the machine learning literature, the output of a model is oftentimes referred to as a prediction. This convention is used in this work, without implying that model estimates are based only on past measurements.

\subsection{Frequency response function model} \label{sec:FRF}
A dynamic system with multiple inputs as well as multiple outputs (MIMO) is characterized by a set of input and output channels. Using $x_k(t)$ to denote a physical quantity measured for the $k$-th input channel at time $t$ and $y_l(t)$ to identify the corresponding measurement for the $l$-th output channel, their relation in the time domain is given by
\begin{align}
	\V y(t) = \M g(t) \ast \V x(t)
\end{align}
with $\M g(t)$ being the weight function matrix. Unfortunately, measuring the components of the weight function matrix directly, by applying an impulse excitation with sufficient energy content at the input channels, would severely damage the system in many practical applications. Moreover, the calculation of the convolution operation $(\ast)$ using Duhamel's integral is very time consuming. It is therefore common to compute the frequency response function matrix $\M H(\imag \omega)$ in the frequency domain
\begin{align}
	\V Y(\imag \omega) = \M H(\imag \omega) \V X(\imag \omega),
\end{align}
where $\V X \left(\imag \omega \right)=\mathcal{F} \left( \V x \left(t \right) \right)$ and $\V Y \left(\imag \omega \right)=\mathcal{F} \left( \V y \left(t \right) \right)$  are the Fourier transforms of the input and output channel vectors. The elements of $\M H (\imag \omega)$ can be estimated as
\begin{align}
	H_{kl} (\imag \omega_n) = \frac{\bar{S}_{kl} (\imag \omega_n)}{\bar{S}_{kk} (\imag \omega_n)}
\end{align}
where the effective power spectral density (PSD) $\bar{S}_{kk}$ and effective cross power spectral density $\bar{S}_{kl}$ can be computed as
\begin{align}
	\bar S_{kk} (\imag \omega_n) &= \frac{1}{T \cdot M} \sum_{m=1}^{M} X_{m,k}^\ast (\imag \omega_n) X_{m,k} (\imag \omega_n) \\
	\bar S_{kl} (\imag \omega_n) &= \frac{1}{T \cdot M} \sum_{m=1}^{M} X_{m,k}^\ast (\imag \omega_n) Y_{m,l} (\imag \omega_n)
\end{align}
by averaging over the $M$ signal windows of size $T$ for each discrete $\omega_n$ obtained from the Fast Fourier Transform (FFT). This procedure allows for a system characterization using a white-pink-noise excitation. For further details, the reader is referred to \cite{Natke1988}, \cite{Dodds2001} and \cite{Hay2007}.\\
After parameterization, the frequency response function matrix can be used as a predictive model which computes output channel signals perfectly from known input channel information if the dynamic system is linear and time invariant. In the single input, single output (SISO) case, this method can also be extended to non-linear systems by a system linearization, see \cite{Hay2007}, or the usage of multiple models at different support points of a piecewise linearization. In the MIMO case, however, the total number of support points increases exponentially as the non-linear characteristic of each channel combination can be different, making the parameterization and application of this approach very challenging. 

\subsection{Windowed LSTM prediction}
The Long Short-Term Memory network, introduced in Hochreiter and Schmidhuber\cite{Hochreiter1997, Gers1999}, is a type of artificial neural network used for sequential data processing. Its gated structure was designed to overcome the vanishing gradient problem of previous recurrent neural networks. As a result, these networks can be trained more efficiently and are generally better at memorizing relationships over long time periods. While many applications of LSTM networks are designed to forecast a time series from data samples of its history in an autoregressive manner, this work applies them to approximate responses of a physical quantity of interest, using concurrent measurements of other physical quantities as input data. This section only provides a short overview of the LSTM network algorithm. For more detailed information, the reader is referred to the aforementioned works.\\
When applied to measurement data from dynamic systems with high sampling rates, processing a long measurement times series in a single LSTM network prediction would lead to significant performance issues. These are overcome by conducting model training and inference using time windows, which are ultimately recombined into a continuous prediction.

\subsubsection{LSTM networks}
An LSTM network is composed of one or multiple memory blocks. Each memory block contains a number of memory cells, where $\V c(t_i)$ denotes the vector of inner cell states at time step $t_i$ with  $i=1\dots L$.
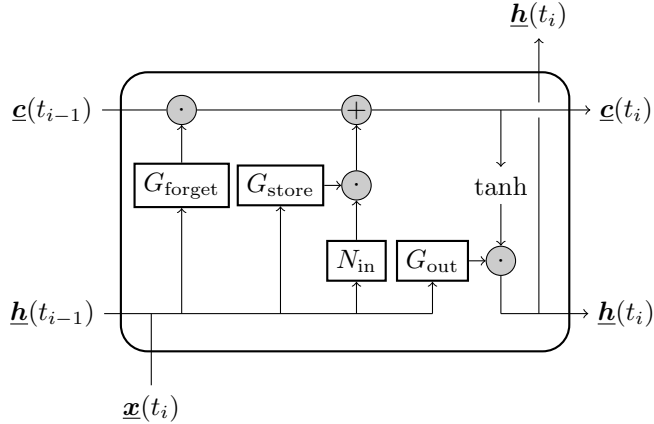
\begin{figure}
	\begin{centering}
\tikzstyle{net}=[rectangle,draw=black!100,fill=black!0, thick]
\tikzstyle{op}=[circle,draw=black!100,fill=black!20, minimum size=11, inner sep=0]

\begin{tikzpicture}

\def\Yt{3.7};  
\def\Ymt{2.7}; 
\def\Ymm{2.7}; 
\def\Ymb{1.7}; 
\def\Yb{1}; 


\def\Xpast{0};
\def\Xci{0.8};
\def\Xf{1.2};
\def\Xs{2.5};
\def\Xi{3.5};
\def\Xo{4.5};
\def\Xt{5.4};
\def\Xu{5.9};
\def\Xnext{6.8};

\node (Bi) at (\Xci, \Yb-1.25) {$\V x (t_i)$};
\node (Bhp) at (\Xpast-0.5, \Yb) {$\V h (t_{i-1})$};
\node (Bcp) at (\Xpast-0.5, \Yt) {$\V c (t_{i-1})$};
\node (Bhb) at (\Xnext+0.25, \Yb) {$\V h (t_i)$};
\node (Bht) at (\Xu,\Yt+1.25) {$\V h (t_i)$};
\node (Bct) at (\Xnext+0.25, \Yt) {$\V c (t_i)$};

\node (Gf) at (\Xf, \Ymm) [net]{$G_\text{forget}$};
\node (Gs) at (\Xs, \Ymm) [net]{$G_\text{store}$};
\node (Ni) at (\Xi, \Ymb) [net]{$N_\text{in}$};
\node (Go) at (\Xo, \Ymb) [net]{$G_\text{out}$};

\node (tanh) at (\Xt, \Ymt) {$\tanh$};

\node (Mf) at (\Xf, \Yt) [op]{$\cdot$};
\node (Ms) at (\Xi, \Ymm) [op]{$\cdot$};
\node (Ai) at (\Xi, \Yt) [op]{$+$};
\node (Mo) at (\Xt, \Ymb) [op]{$\cdot$};

\draw[->] (Bcp) -- (Mf) -- (Ai) -- (Bct);

\draw (Bhp) -- (\Xo,\Yb);
\draw[->] (\Xt,\Yb) -- (Bhb);

\draw (Bi) -- (\Xci,\Yb);

\draw[->] (\Xf,\Yb) -- (Gf);
\draw[->] (Gf) -- (Mf);

\draw[->] (\Xs,\Yb) -- (Gs);
\draw[->] (Gs) -- (Ms);
\draw[->] (\Xi,\Yb) -- (Ni);
\draw[->] (Ni) -- (Ms);
\draw[->] (Ms) -- (Ai);

\draw[->] (\Xt,\Yt) -- (tanh);
\draw[->] (tanh) -- (Mo);
\draw (Mo) -- (\Xt,\Yb);
\draw[->] (\Xo, \Yb) -- (Go);
\draw[->] (Go) -- (Mo);

\draw (\Xu,\Yb) -- (\Xu,\Yt-0.1);
\draw[->] (\Xu,\Yt+0.1) -- (Bht);

\draw[draw=black, rounded corners=10, thick] (\Xf-0.8, \Yb-0.5) rectangle ++(\Xu-\Xf+1.2,\Yt);

\end{tikzpicture}
		\caption[LSTM setup]{LSTM networks are composed of memory blocks, which process information using the inner cell state $\V c$. This inner state can be updated, forgotten or used to create the block output $\V h$ depending on the block input at the current time step $\V x (t_i)$ and the previous output $\V h (t_{i-1})$. All operations related to the inner state are carried out by respective gates $G$ and the network $N$, whose parameters are learned from a dataset during training.}
		\label{fig:lstm}
	\end{centering}
\end{figure}
\autoref{fig:lstm} provides an overview of the data processing steps that occur in a single memory block. The inner state of each memory cell can be updated, forgotten or used to create the block output $\V h$ depending on the block input at the current time step $\V x (t_i)$ and the previous output $\V h (t_{i-1})$. All operations related to the inner state are carried out by respective gates $G$ and the network $N$, whose parameters are learned from a dataset during training.\\
At the beginning of each prediction, the cell states are initialized to zero. The LSTM model processes the input channel values $\V x (t_i)$ in sequence at discrete points in time, starting at $\V x(t_1)$ and incrementing $i$ until the terminal input $\V x (t_L)$ of the sequence with length $L$ is reached. The block input $\V x (t_i)$ is concatenated with the output of the previous timestep $\V h(t_{i-1})$ to provide input data for the gate networks $G_\text{store}$, $G_\text{forget}$, $G_\text{out}$ and the input network $N_\text{in}$. Each gate network uses the sigmoid activation function $\sigma(x) = 1/(1+\e^{-x})$ and provides an output
\begin{alignat}{3}
	\V g_\text{store} (t_i)   &= \sigma \big(  \M W_\text{store}^x \V x (t_i)  & &+ \M W_\text{store}^h \V h (t_{i-1})  & &+ \V b_\text{store} \big) \\
	\V g_\text{out} (t_i)     &= \sigma \big(  \M W_\text{out}^x \V x (t_i)    & &+ \M W_\text{out}^h \V h (t_{i-1})    & &+ \V b_\text{out} \big) \\
	\V g_\text{forget} (t_i)  &= \sigma \big(  \M W_\text{forget}^x \V x (t_i) & &+ \M W_\text{forget}^h \V h (t_{i-1}) & &+ \V b_\text{forget} \big)
\end{alignat}
of the respective gate in the range (0,1), while the input network
\begin{align}
	\V a_\text{in} (t_i) = \tanh \left(  \M W_\text{in}^x \V x (t_i) + \M W_\text{in}^h \V h (t_{i-1}) + \V b_\text{in} \right)
\end{align}
uses the tanh activation function. The parameters of each network are given by the weight matrices $\M W$ and the bias vector $\V b$, which are initialized randomly. The memory cell state of the current time step
\begin{align}
	\V c (t_i) = \V c (t_{i-1}) \odot \V g_\text{forget} (t_i) + \V a_\text{in} (t_i) \odot \V g_\text{store} (t_i)
\end{align}
as well as the block output
\begin{align}
	\V h \left(t_i\right) = \tanh \left( \V c \left(t_i\right) \right) \odot \V g_\text{out} \left(t_i\right)
\end{align}
are now computed using the Hadamard product $\odot$ for element-wise multiplication. Afterwards, this process is repeated for the next time step. In architectures with multiple LSTM blocks, the output vector $\V h$ of one block is used as the input $\V x$ of the next block. After the final block, a single fully connected layer
\begin{align}
	\V y^\ast (t_i) = \M W_\text{FC} \V h (t_i) + \V b_\text{FC}
\end{align}
with weights $W_\text{FC}$, bias $b_\text{FC}$ and no activation function is used to generate the network prediction $\V y^\ast$.\\
In this work, the LSTM network is implemented using the Python libraries Tensorflow \cite{tensorflow} and Keras \cite{keras}. As a preprocessing step, all training input and output data is collectively standardized to a mean of zero and a standard deviation of one in each channel. During training, the input sequence is used to generate a prediction
\begin{align}
	\V y^\ast \left(t_i\right) = \text{LSTM} \left( \V x \left(t_i\right) \right),
\end{align}
which is then compared to the true output channel values using the mean squared error loss function
\begin{align}
	E_\text{MSE} = \frac{1}{L} \sum_{i=1}^{L} \left( \V y \left(t_i\right) - \V y^\ast \left(t_i\right) \right)^2.
\end{align}
The optimization algorithm RMSProp \cite{RMSProp} updates the weights and biases after each mini-batch of training data in order to minimize the loss function. The rate of change of these updates is determined by the learning rate hyper-parameter $\lambda$. Training proceeds for a given number of training dataset repetitions called epochs and the assignment of data samples to mini-batches is randomized after each epoch.\\

\subsubsection{Windowing}
\begin{figure*}
	\begin{center}
		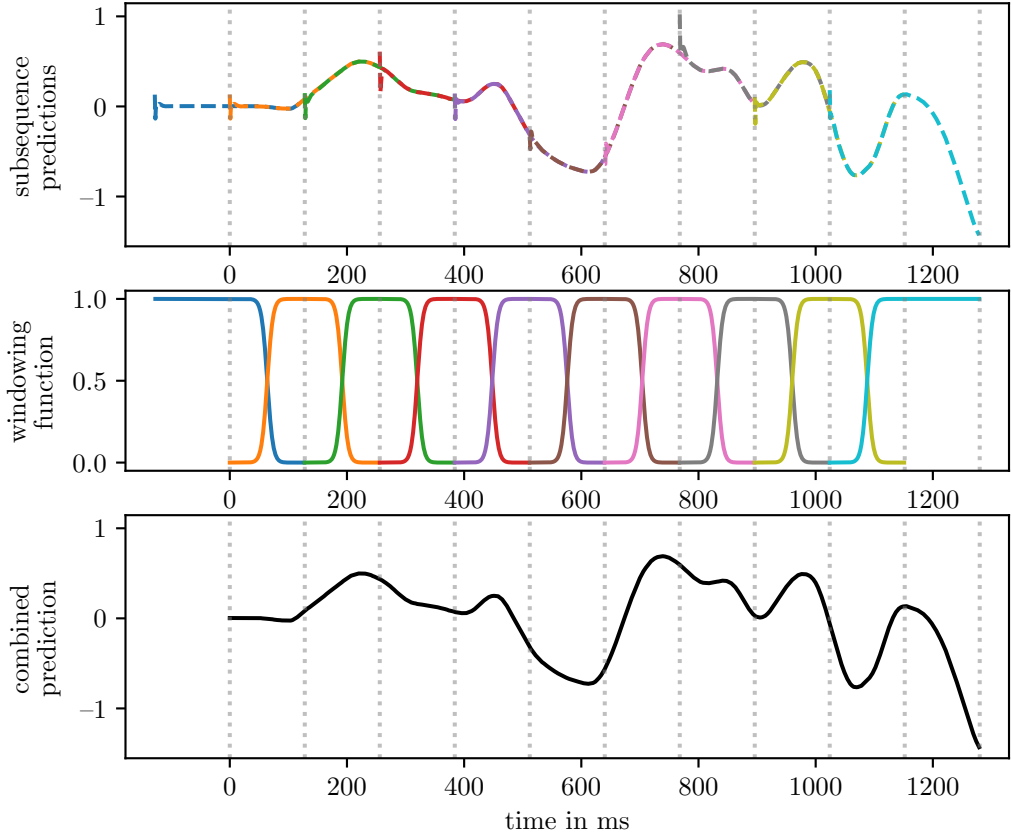
		\caption[Windowed prediction]{Individual subsequence predictions (top) are combined as a weighted sum using corresponding window functions (center). As a consequence, predictions for long sequences (bottom) resulting from measurement data can be realized.}
		\label{fig:windowed_prediction}
	\end{center}
\end{figure*}

Time series processed by LSTM networks can generally vary in their number of samples and duration. However, long sequences require the sequential computation of gradients for weights and biases over many time steps, which is computationally expensive and inefficient. Furthermore, the dynamic response of a damped mechanical system subject to dynamic loads depends only on the current loads and a limited loading history, since all excited oscillations decay after a certain time. As a result, the ability of LSTM networks to memorize information over very long periods of time is less important in the context of fatigue testing.\\
The application of LSTM networks to measurement data from sensors is therefore carried out on short subsequences with a fixed length $L$. The extraction of subsequences from the dataset is rather straightforward. Starting at the first time sample $t_i = t_1$ of each measurement data file, data in the range $t \in [t_i, t_{i+L-1}]$ is extracted and the starting position of the next subsequence is given by
\begin{align}
	t_i \leftarrow t_i + o \cdot L,
\end{align}
where the overlap factor $o$ determines the number of shared time samples between neighboring subsequences.\\
Both training and prediction of the LSTM network are carried out on the subsequence level. In order to achieve a continuous prediction $y^\ast_\text{comb}$ for a complete measurement file, the $n_\text{sub}$ individual subsequence predictions $y^\ast_\text{sub}$ are combined as a weighted sum
\begin{align}
	\V y^\ast_\text{comb} = \sum^{n_\text{sub}} \V y^\ast_\text{sub} \odot \V w_\text{sub} \label{eq:weighting}
\end{align}
using window functions $w_\text{sub}$, which are non-zero only in the range of the corresponding subsequence. The window functions are also weighted to ensure the partition of unity property
\begin{align}
	\sum^{n_\text{sub}} \V w_\text{sub} = 1
\end{align}
is fulfilled at every time step. \autoref{fig:windowed_prediction} illustrates the subsequence combination process for a single output channel.
\\
In this paper, the Welch window function
\begin{align}
	w_\text{sub} \left( t \right) = 1 - \left( \frac{2t - L}{L} \right)^2
\end{align}
is used to interpolate between individual predictions. It is additionally raised to the power of 10 to provide smooth transitions. As a result of this weighting scheme, the first and last 15\% of the data in each subsequence have a very low impact on the combined prediction. This is essential, since error levels are increased for predictions at the start of each window, as information on the recent history before the beginning of the subsequence is lacking. A special treatment is used to extend this approach to the beginning of the measurement process. Here, an additional subsequence is utilized, which spans from $t=-L/2$ to $t=L/2$. For all negative points in time, the unmeasured input channel data is assumed to be equal to the corresponding measurements at the starting time $t_1$. A prediction is generated for this subsequence and included into the weighting process. Finally, all negative points in time are discarded from the resulting combined prediction.

\subsection{Hybrid modeling strategies} \label{sec:hybrid_models}
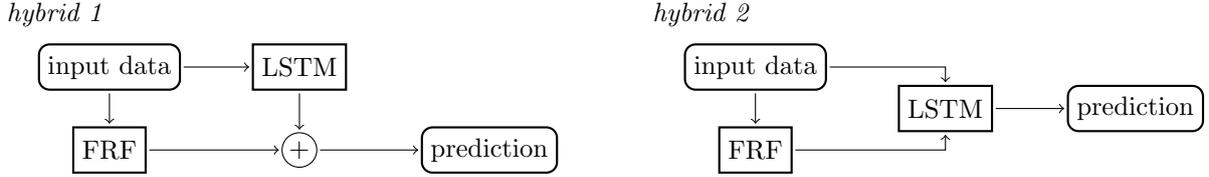
\begin{figure*}
	\begin{center}
%
%
%
%

\begin{tikzpicture}
	
	\def\Scale{1.0};
	\def\XInput{0};
	\def\YInput{0};
	\def\XTitle{0.7*\Scale};
	\def\YTitle{0.7*\Scale};
	\def\XLSTM{2.5*\Scale};
	\def\YFRF{1.1*\Scale};
	\def\XHybridOne{2.5*\Scale};
	
	\node (TitleOne) at (\XInput-\XTitle, \YInput+\YTitle) {\textit{hybrid 1}};
	\node[draw, thick, align=left, rounded corners, minimum height=0.6cm] (InDataOne) at (\XInput, \YInput) {input data};
	\node[draw, thick, align=left, minimum height=0.6cm] (LSTMone) at (\XInput+\XLSTM, \YInput) {LSTM};
	\node[draw, thick, align=left, minimum height=0.6cm] (FRFone) at (\XInput, \YInput - \YFRF) {FRF};
	\node[circle, draw=black!100,fill=gray!0, minimum size=13, inner sep=0] (PlusOne) at (\XInput+\XLSTM, \YInput - \YFRF) {+};
	\node[draw, thick, align=left, rounded corners, minimum height=0.6cm] (HybridOne) at (\XInput+\XLSTM+\XHybridOne, \YInput - \YFRF) {prediction};
	
	\begin{scope}[->,shorten >=1pt, shorten <=1pt]
		
	\draw[->] (InDataOne) -- (LSTMone);
	\draw[->] (InDataOne) -- (FRFone);
	\draw[->] (LSTMone) -- (PlusOne);
	\draw[->] (FRFone) -- (PlusOne);
	\draw[->] (PlusOne) -- (HybridOne);
	
	\end{scope}

	\def\XSchemeTwo{8.5*\Scale};
	\def\YFRFTwo{1.1*\Scale};
	\def\XLSTM{2.5*\Scale};
	\def\XHybridTwo{2.5*\Scale};
	
	\node (TitleTwo) at (\XInput+\XSchemeTwo-\XTitle, \YInput+\YTitle) {\textit{hybrid 2}};
	\node[draw, thick, align=left, rounded corners, minimum height=0.6cm] (InDataTwo) at (\XInput+\XSchemeTwo, \YInput) {input data};
	\node[draw, thick, align=left, minimum height=0.6cm] (FRFtwo) at (\XInput+\XSchemeTwo, \YInput-\YFRFTwo) {FRF};
	\node[draw, thick, align=left, minimum height=0.6cm] (LSTMtwo) at (\XInput+\XSchemeTwo+\XLSTM, \YInput -0.5*\YFRFTwo) {LSTM};
	\node[draw, thick, align=left, rounded corners, minimum height=0.6cm] (HybridTwo) at (\XInput+\XSchemeTwo+\XLSTM+\XHybridTwo, \YInput -0.5*\YFRFTwo) {prediction};
	
	\begin{scope}[->,shorten >=1pt, shorten <=1pt]
	
	\draw[->] (InDataTwo) -- (FRFtwo);
	\draw[->] (InDataTwo) -- (\XInput+\XSchemeTwo+\XLSTM, \YInput) -- (LSTMtwo);
	\draw[->] (FRFtwo) -- (\XInput+\XSchemeTwo+\XLSTM, \YInput-\YFRFTwo) -- (LSTMtwo);
	\draw[->] (LSTMtwo) -- (HybridTwo);
	
	\end{scope}
	
	%
	%
	
\end{tikzpicture}

		\caption[Hybrid schemes]{The hybrid modelling scheme \textit{hybrid 1} uses an LSTM network to apply a nonlinear correction of the linear FRF model prediction. In \textit{hybrid 2}, the LSTM network is instead provided with both the original input data and the corresponding FRF model output, so that the linear solution can be incorporated into the prediction.}
		\label{fig:hybrid_schemes}
	\end{center}
\end{figure*}

The FRF model excels at predicting linear system behavior, while the LSTM network can be trained to approximate arbitrary non-linear time-dependent functions. The aim of a hybrid model is to combine both methods in order to exploit their respective strength while minimizing their downsides. In this paper, two different approaches to hybrid modeling are compared, as shown in \autoref{fig:hybrid_schemes}.\\
\noindent
In the first approach, denoted by \textit{hybrid~1}, the LSTM network is used to improve the FRF model prediction by providing a non-linear correction. The FRF model initially generates a prediction from the input channel values $x$. This prediction is then subtracted from the true output channel values $y$ in the dataset
\begin{align}
	\V e \left( t\right) = \V y \left( t \right) - \text{FRF} \left( \V x \left( t \right) \right)
\end{align}
in order to obtain the model error $e$ of the linear FRF prediction. This error data can now be used as output channel values during the training process of an LSTM network. Since the trained network now provides an estimate
\begin{align}
	\V e^\ast \left( t\right) = \text{LSTM} \left( \V x \left( t\right) \right)
\end{align}
for the FRF model error, the sum
\begin{align}
	\V y^\ast_1 \left( t \right) = \text{FRF} \left( \V x \left( t\right) \right) + \text{LSTM} \left( \V x \left( t \right) \right)
\end{align}
yields the hybrid prediction of the model \textit{hybrid~1}.\\
The second approach, denoted by \textit{hybrid~2}, is designed to reduce the learning complexity of the LSTM network by providing it with more information. Again, the FRF model is used to create a prediction, which is now concatenated to the original input data during LSTM training. This way, the training dataset already contains the FRF prediction as a baseline solution. The hybrid prediction of \textit{hybrid~2}
\begin{align}
	\V y^\ast_2 \left( t \right) = \text{LSTM} \left( \V x \left( t \right), \text{FRF} \left( \V x \left( t\right) \right) \right)
\end{align}
can now simply be a copy of the FRF prediction or a modification by the LSTM, depending on the input channel data $x$.

\section{Experiments} \label{sec:experiments}
Forward prediction (FP) can achieve substantial benefits in experimental fatigue tests under multiaxial variable amplitude loading. Virtual sensing (VS) provides accurate sensor signal predictions from easily accessible measurements, which is important for enabling fatigue monitoring and predictive maintenance approaches. Therefore, a real live measurement setup of a fatigue test is chosen in this work to provide a realistic example for the FP use case. In VS applications, the system excitation is typically unknown and the simulation model estimates a relationship between multiple system responses. As such, this dataset does not represent a realistic application example of VS, but rather an exemplary non-linear system where the relationship between system responses can be identified. \autoref{fig:VSFP} shows how the fatigue test bench setup is used for FP and VS signal estimation tasks in this work. The measurement data of this paper is made publicly available \cite{OpARA} and can be accessed via the DOI \href{http://dx.doi.org/10.25532/OPARA-151}{\textit{http://dx.doi.org/10.25532/OPARA-151}} to provide a reproducible and large dataset for further research.
\begin{figure}
	\begin{centering}
\begin{tikzpicture}

\node[draw, thick, align=left, rounded corners, minimum height=0.6cm](Input) at (0.05,0) {drive signal};
\node[draw, thick, align=left, minimum height=0.6cm](System) at (2.7,0) {test bench};
\node[draw, thick, align=left, rounded corners, minimum height=0.6cm](Output) at (5.5,0) {measurement};

\node (frc1) at (-0.4, -0.75) {$\bullet$ force};

\node (dsp2) at (5.53, -0.75) {$\bullet$ displacement};
\node [below=0.5 cm of dsp2.west,anchor=west] (frc2)  {$\bullet$ force};

\begin{scope}[->,shorten >=2pt, shorten <=2pt]
\draw [->] (Input) -- (System);
\draw [->] (System) -- (Output);

\draw [draw=orange, ->, thick] (frc1) -- (dsp2);
\draw [draw=orange, ->, thick, rounded corners=4pt] (frc1) -- (2.5,-0.75) -- (2.5,-1.25) -- (frc2);
\draw [draw=blue, ->, thick, rounded corners=4pt] (dsp2) -- (7,-0.75) -- (7,-1.25) -- (frc2);
\end{scope}

\node[text=orange, thick] at (1.3,-1.03) {FP};
\node[text=blue, thick] at (7.4,-1.03) {VS};

\end{tikzpicture}
		\caption[Prediction modes]{The proposed hybrid models can be applied to different signal estimation problems. In forward prediction (FP), the output sensor data is estimated based on the drive signal, which controls the system excitation. In a virtual sensing task (VS), one or more output sensors are estimated from the remaining measurements. The assignment of input and output data during model parameterization changes depending on the use case.}
		\label{fig:VSFP}
	\end{centering}
\end{figure}
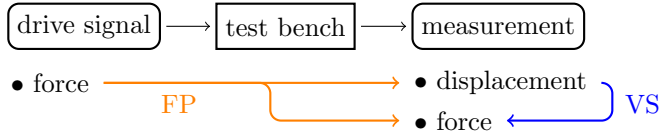\\
\subsection{Experimental setup and data} \label{sec:experimental_setup}
\begin{figure*}
	\begin{center}
		\subfloat[Fatigue test bench]{
			\includegraphics[height=0.34\textwidth]{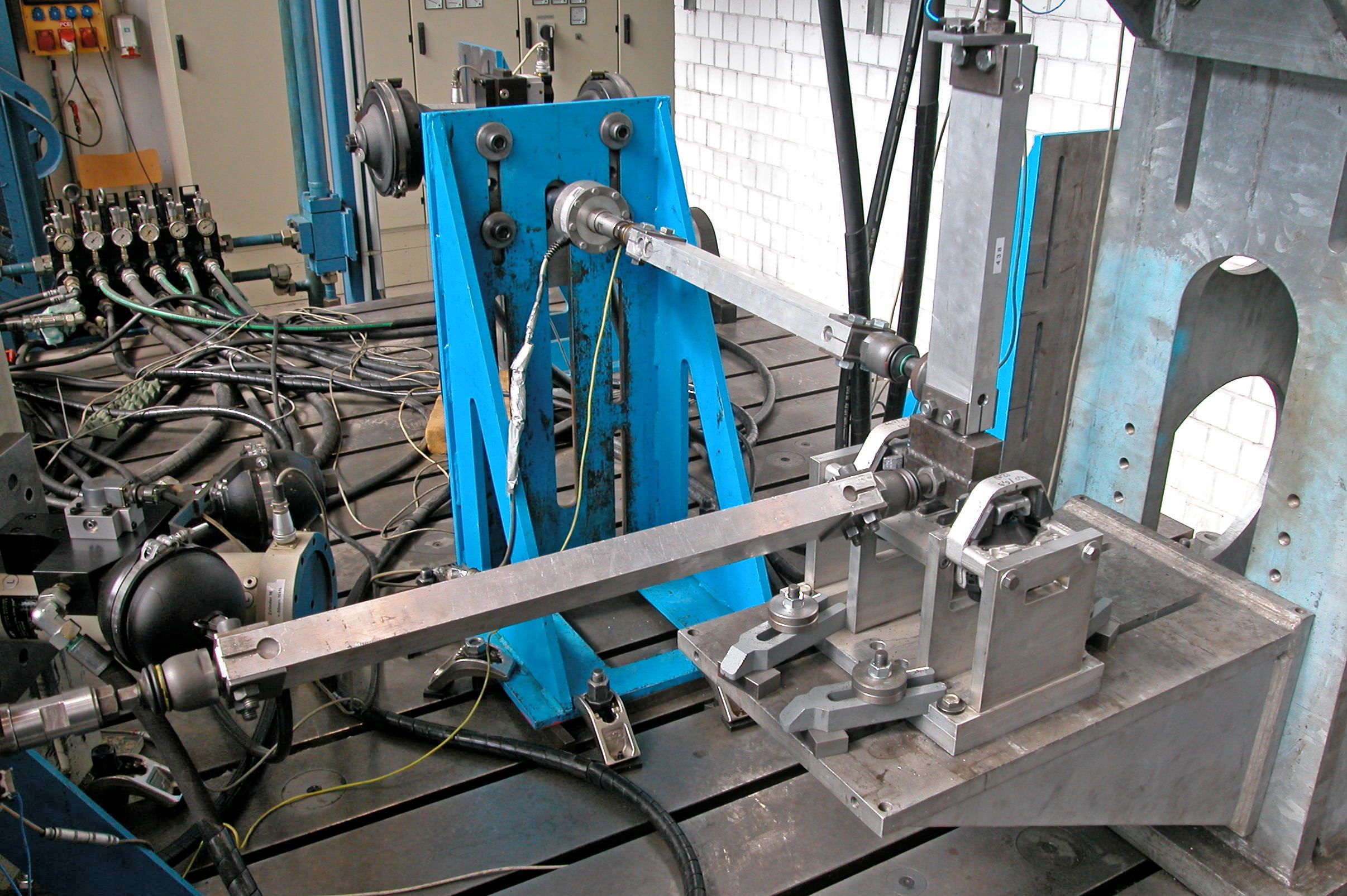}
		}
		\subfloat[Hydro-mount]{
			\includegraphics[height=0.34\textwidth]{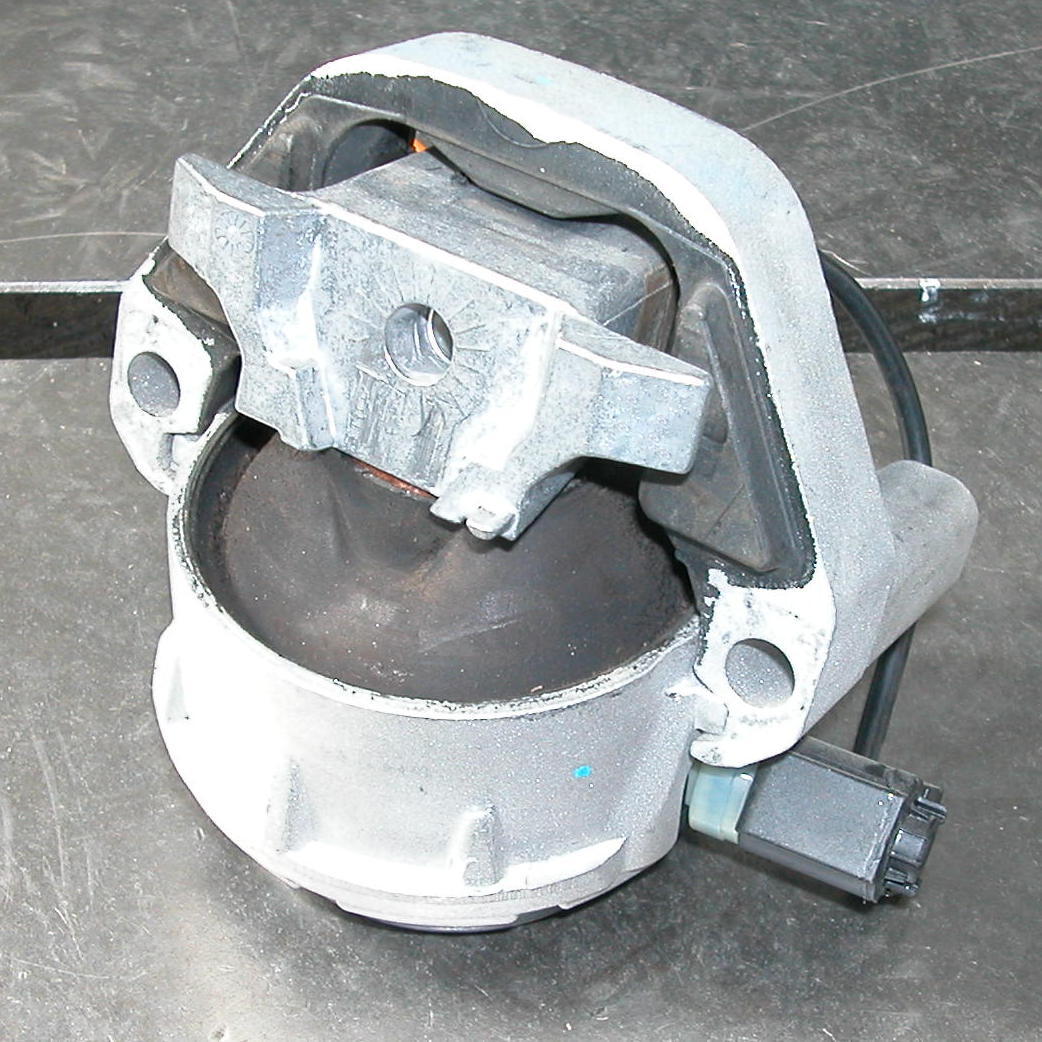}
		}
		\caption[Test setup]{A three-component servo-hydraulic test bench (a) for the fatigue assessment of suspension hydro-mounts (b) is used to generate an experimental dataset.}
		\label{fig:test_setup}
	\end{center}
\end{figure*}

\begin{figure*}
	\begin{center}
		\subfloat{
			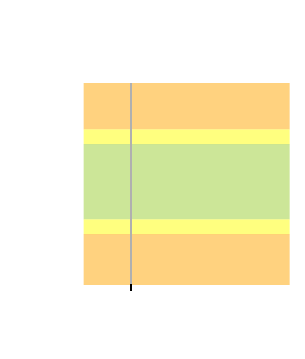
		}
		\subfloat{
			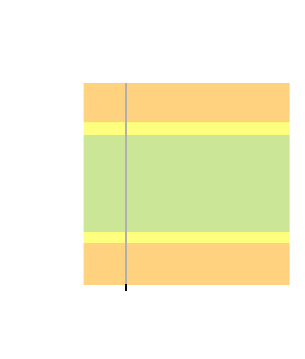
		}
		\subfloat{
			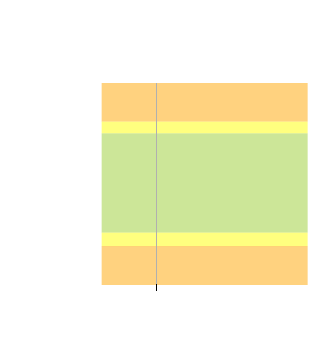
		}
		\caption[Static characteristics]{The static characteristics of each spatial direction form a friction induced hysteresis. While the stiffness is almost constant in large regions of the parameter space, depicted in green color, it changes significantly in the vicinity of the lower and upper reversal points as indicated by the yellow and orange areas. Only data from regions of constant stiffness is used in the parameterization of the FRF model.}
		\label{fig:non-linear_characteristics}
	\end{center}
\end{figure*}

\begin{figure*}
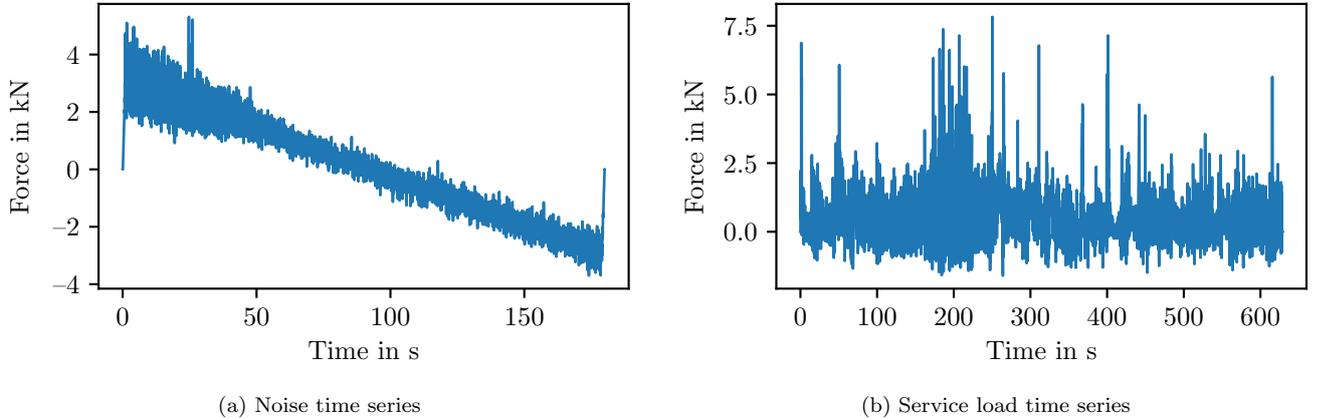

	\begin{center}
		\subfloat[Noise time series]{
			\input{Noise_plot_thin.pdf_tex}
		}
		\subfloat[Service load time series]{
			\input{Serviceload_plot_thin.pdf_tex}
		}
		\caption[Example time series]{The experimental dataset contains a large amount of measured noise time series (a), which are used to train LSTM networks and parameterize FRF models. For validation and testing, mainly service load data files (b) are used to emulate fatigue testing conditions.}
		\label{fig:example_time_series}
	\end{center}
\end{figure*}

In order to demonstrate and validate the proposed approach, experimental data is collected from the three-component servo-hydraulic fatigue test bench for suspension hydro-mounts, depicted in \autoref{fig:test_setup}. It is equipped with 3 inertia compensated force and 3 displacement sensors.
This system features a variety of non-linearities. The hydro-mounts are filled with oil to provide highly non-linear damping, while the pendulum kinematics of the setup introduce non-linear interactions of the excitations in different spatial directions. The most influential non-linearity originates from the system stiffness. While the static force-displacement relationship does not sufficiently describe the system dynamics and the directional coupling between the axes, it still provides insights into the nature of the non-linearity. In order to capture it, each channel is loaded and measured individually in a quasi-static manner, while the two remaining channels are load controlled at a force of zero. The resulting static characteristics is depicted in \autoref{fig:non-linear_characteristics}.
Apart from the hysteresis, the stiffness is nearly constant in the central regions of parameter space. The FRF model is parameterized using only noise data from this region in order to provide the best linear approximation to the overall system behavior.\\
The measured dataset contains a large collection of system responses that result from different excitations, which are sampled with a frequency of \SI{1}{\kilo\hertz}. It can be subdivided into \SI{1}{\hour} \SI{53}{\minute} of uncorrelated noise signals, \SI{2}{\hour} \SI{37}{\minute} of fatigue service loads whose input signals are rescaled from three independent service load shapes, \SI{20}{\minute} of sinusoidal excitations and \SI{20}{\minute} of sweep. \autoref{fig:example_time_series} shows a time series example for noise and service load data, respectively.
In order to ensure that the system characteristics are constant over the duration of all measurements, the entire testing program was repeated once, resulting in identical responses according to the sensor accuracy. A subsequently performed data postprocessing only included a low pass FFT-filtering at \SI{80}{\hertz}. The remaining signal contains the complete controllable frequency range of the used test setup.\\
For the purpose of LSTM network parameterization, the dataset is split into three parts, namely training, validation and test data. Only the training dataset is directly used to update the network weights, while the validation dataset is used to determine the best choice of hyper-parameters of the LSTM architecture and subsequence windowing process. The test dataset is never used during parameterization in order to enable a completely independent evaluation of the predictive performance for each model.\\
From a fatigue analysis point of view, it is very desirable to parameterize predictive models using noise signals exclusively, as they contain a lot of information on the system, while causing significantly less damage to the test specimen if compared to service loads. For this reason, the majority of the noise data is used for training while only a small portion is left to asses the noise prediction quality during validation and testing. The service load data on the other hand is only used during validation and testing. Here, two of the three basic service load signal shapes are assigned for validation and the remaining one is reserved for testing to ensure that both datasets are completely independent. The composition of the dataset is specified in more detail in \autoref{tab:dataset}.
\begin{table*}
	\centering
	\caption[Dataset composition]{Composition of the experimental dataset, sampled at \SI{1}{\kilo \hertz}}
	\label{tab:dataset}
	\begin{tabular}{ p{2.5cm} R{2.5cm} R{2cm} R{2cm} R{2cm} }
		\toprule
		Data type & Runtime & \multicolumn{3}{c}{Proportional usage} \\
		& (total) &  Training & Validation & Test \\	
		\toprule
		Noise & \SI{1}{\hour} \SI{53}{\minute} & 85\% & 6\% & 9\% \\
		Service load & \SI{2}{\hour} \SI{37}{\minute} & 0\% & 72\% & 28\% \\
		Sinusoidal & \SI{20}{\minute}  & 0\% & 80\% & 20\%\\
		Sweep & \SI{20}{\minute} & 0\% & 80\% & 20\% \\
		\bottomrule
	\end{tabular}
\end{table*}\\
It should be noted that fatigue tests can generally result in degradation of the test specimen, which would influence the observed system behavior. These effects are not considered in this paper, since the presented virtual sensing approach is targeted at predictive maintenance applications, where the sensor data is used to preemptively replace components before degradation occurs.

\subsection{Error metrics}
In order to evaluate the prediction quality of the models, a variety of error metrics are taken into account. The Root Mean Square (RMS) error
\begin{align}
	\text{RMS}\left(y^\ast, y\right) = \sqrt{\frac{\sum_{l=1}^{T} \left(y \left(t_l\right)-y^\ast \left(t_l\right)\right)^2}{\sum_{l=1}^{T} y\left(t_l\right)^2}}
\end{align}
provides a general measure of how well the signal shape of a particular channel prediction $y^\ast$ corresponds to the target $y$, averaged over all $T$ time steps of the series. Similarly, the power spectral density RMS error
\begin{align}
	\text{RMS}_\text{PSD} \left( y^\ast, y \right) = \text{RMS} \left( S_{kk}\left( y^\ast \right), S_{kk}\left( y \right) \right)
\end{align}
uses the PSD $S_{kk}$ introduced in \autoref{sec:FRF} to evaluate the prediction quality of a signal channel in the frequency domain, averaged over the corresponding frequencies $\omega_n$.\\
Both of these error metrics do not provide information about the approximation quality of the signals' global extrema, which are of high importance for accurate fatigue damage predictions. For this reason, fictitious fatigue damage calculations according to the nominal stress concept \cite{Haibach2002} are performed for both prediction and target signal. Here, a fictitious Wöhler curve
\begin{align}
	N = K \cdot S_\text{a}^{-k}
\end{align}
relates between the load amplitude $S_\text{a}$ and the number of load cycles before component failure denoted by $N$, with $k=5$ and $K=10^7$. In addition, the 4-point Rainflow counting algorithm \cite{McInnes2008} and the elementary Palmgren-Miner rule \cite{Palmgren1924} are used to calculate a fictitious accumulated fatigue damage $d$ for the prediction and target signal, respectively. The metric of the damage ratio
\begin{align}
	\text{damage}(y^\ast, y) = \frac{d (y^\ast)}{d (y)}
\end{align}
therefore informs about the relative error between predicted and target fatigue damage, introduced by the virtual sensing model. 
The Multi-Rain fatigue damage generalizes the fatigue damage accumulation to multiple spatial directions, see Beste et al. \cite{Beste1992}. This necessity arises as a result of multiaxial stress states following from multiaxial component loading.  For a given direction $\psi$, specified by the components $\psi_x$, $\psi_y$ and $\psi_z$ of its unit direction vector satisfying
\begin{align}
	\psi_x^2 + \psi_y^2 + \psi_z^2 = 1,
\end{align}
the fictitious damage in this direction can be computed using a weighted sum
\begin{align}
	d_\psi (\V s) = d (\psi_x s_x + \psi_y s_y + \psi_z s_z)
\end{align}
of the original signal channels $s_x$, $s_y$ and $s_z$ for arbitrary signals $\V s$. The Multi-Rain damage ratio
\begin{align}
	\text{damage}_\text{MR} \left( \V y^\ast, \V y \right) = \frac{\max \left( d_\psi \left( \V y^\ast \right) \right)}{\max \left( d_\psi \left( \V y \right) \right)}
\end{align}
follows by comparing the maximum damages of prediction and target, where 500 uniformly distributed spatial directions $\psi$ are considered in each case.

\subsection{Forward prediction evaluation} \label{sec:forward_prediction_evaluation}
\begin{figure*}
	\begin{center}
		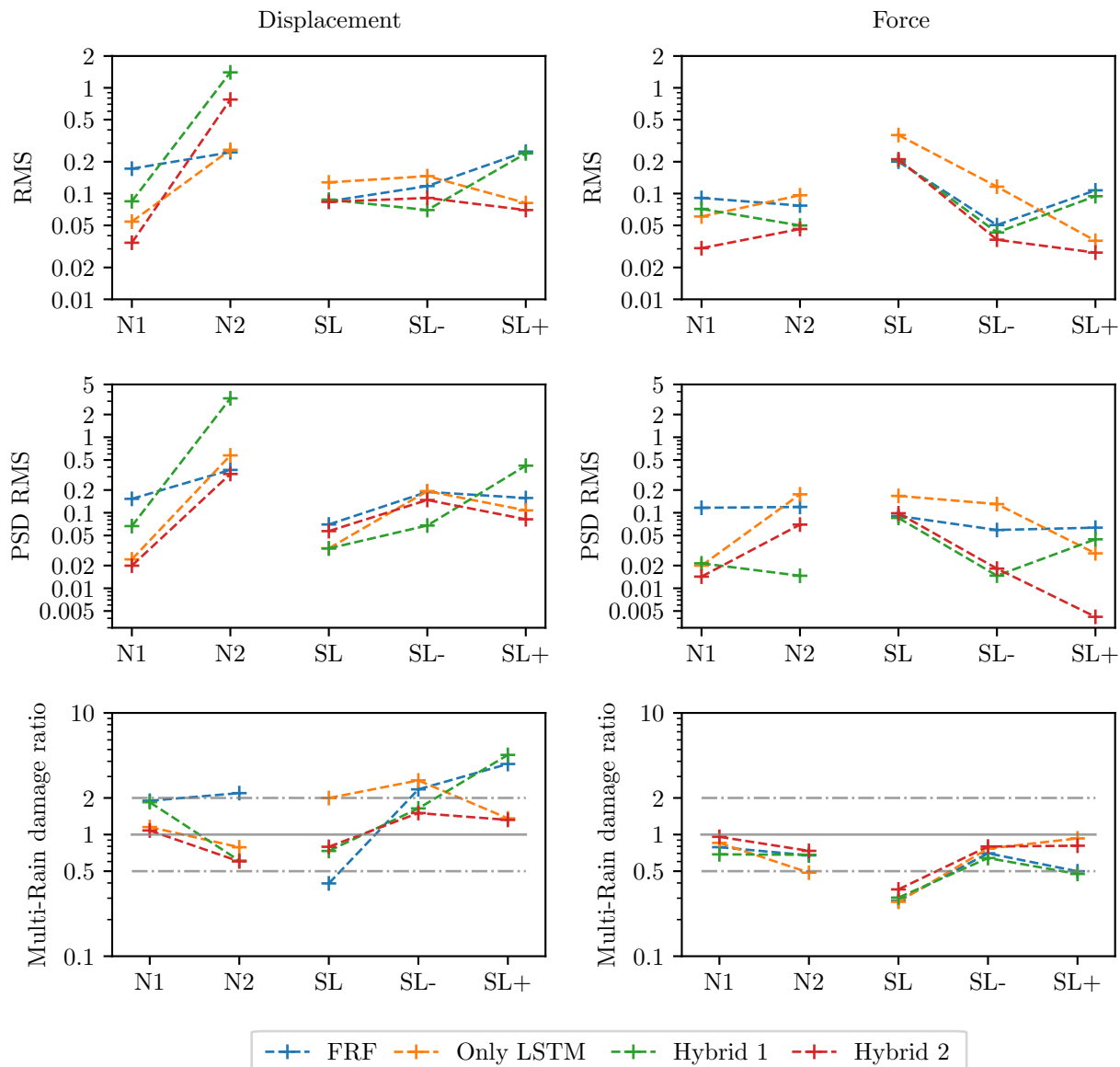
		\caption[Evaluation of forward prediction approach]{In the forward prediction example, both displacement (left) and force (right) sensor data are predicted from the drive signal of the force-controlled servo-hydraulic test bench. FRF, LSTM and hybrid models are compared on all service load and noise test data files. SL shows the average prediction results for the testing service load shape, whose corresponding drive signal is rescaled by factors between 0.3 and 1. SL- and SL+ denote prediction results for the same service load shape with a scaling factor of 0.5 and an additional negative or positive offset applied to the drive signal, which results in highly non-linear system behavior. N1 and N2 represent the prediction results for the two noise measurements in the testing dataset. N1 contains mixed noise with varying degrees of non-linearity, while N2 only includes noise from non-linear boundary regions of the system behavior which are not represented in the training dataset. Both RMS and PSD RMS errors are averaged over the predicted channels.}
		\label{fig:FP_results}
	\end{center}
\end{figure*}

In the forward prediction task, the aim is to predict the measurements of all displacement and force sensors from the test bench drive signal, which determines the system excitation. Using the experimental dataset, a comparison is drawn between the FRF model described in \autoref{sec:FRF}, a pure LSTM network and both hybrid models introduced in \autoref{sec:hybrid_models}. As noted in \autoref{sec:experimental_setup}, the FRF model is parameterized using only data from regions where the system stiffness is nearly constant, while the complete training dataset was used for the LSTM and hybrid models. The network hyper-parameters were chosen after conducting multiple large scale automated parameter studies on a high performance computing cluster. Due to the dataset size, a global optimization in this hyper-parameter space is not feasible. To account for the random initialization process of network parameters, each model architecture was trained three times using different starting initializations. This parameter identification process results in varying learning rates and training epoch numbers of the different approaches. It does not limit the comparability of the respective methods, since, in each category, the model with the best prediction quality of the validation dataset was selected.\\
For all model types, good results were achieved by setting the subsequence length $L$ to 256 and the overlap factor $o$ to 0.5. The best pure LSTM model uses a single LSTM block with 39 memory cells and was trained for 253 epochs with a learning rate $\lambda$ of 0.0002. For the chosen \textit{hybrid~1} model, two memory blocks of 25 memory cells each were trained for 75 epochs using a learning rate $\lambda$ of 0.003. The best \textit{hybrid~2} model again uses one memory block with 39 memory cells and was trained for 800 epochs with $\lambda=0.0001$.\\
The test dataset results for the prediction of noise and service load signal shapes are visualized in \autoref{fig:FP_results}. Predictions of displacements and forces are shown separately, although in all cases both quantities were predicted by the same model. Differently scaled versions of the same drive signal shape were used to generate the testing service loads. The system stiffness changes significantly when an offset is used, resulting in an increased non-linearity of the system behavior.
\\
The results for the testing noise file N1 show, that the LSTM based models significantly outperform the FRF model in all metrics. This is to be expected, since N1 contains mixed noise signals from various regions of the system behavior, which are overall well covered in the LSTM training data. The FRF model performs much worse, since its parameterization data is limited. Noise file N2 includes noise data from a highly non-linear boundary region of the system behavior, which was not included in the LSTM training data. As a result, the performance of LSTM based models deteriorates, since these approaches generally suffer from poor extrapolation. The linear approximation of the FRF model for N2 shows an overall similar result to N1. The service loads without offset, SL, are best predicted by the \textit{hybrid~2} approach across all metrics except PSD RMS for displacement predictions, where it still scores very good results. The \textit{hybrid~2} model also produces the best results for the offset service loads with higher non-linearity, SL- and SL+. Overall, this approach scores closest to the perfect Multi-Rain damage ration of 1 for all service load predictions and for the mixed testing noise N1. From the FP results as a whole, it is evident that both FRF model and pure LSTM network perform very inconsistently by themselves, and that the \textit{hybrid~2} strategy manages to combine their respective strengths very well.\\
\subsection{Virtual sensing evaluation} \label{sec:virtual_sensing_evaluation}
\begin{figure}
	\begin{center}
		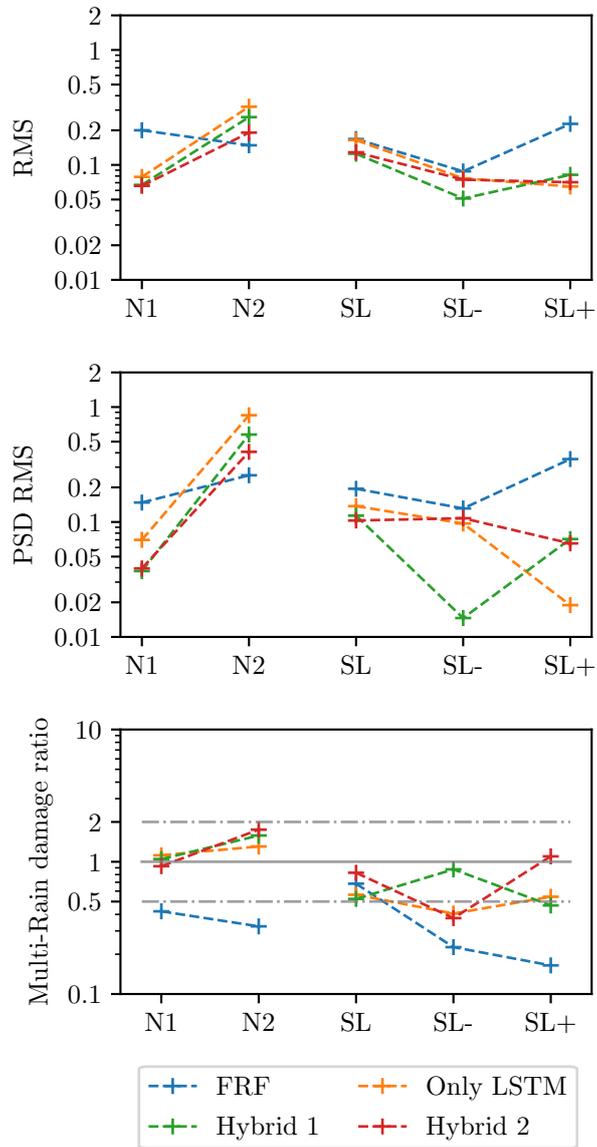
		\caption[Evaluation of virtual sensing approach]{In the virtual sensing example, the force sensor data is predicted using the displacement sensors. The results are structured as described in \autoref{fig:FP_results}.}
		\label{fig:VS_results}
	\end{center}
\end{figure}

\begin{figure}
	\begin{center}
		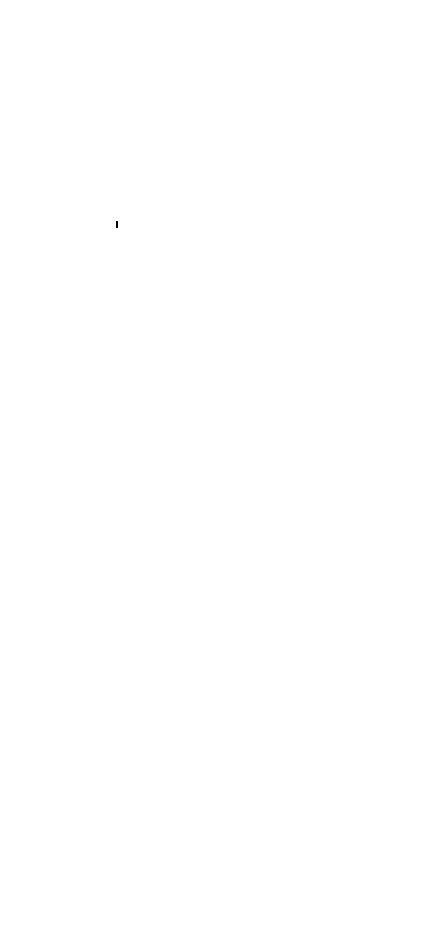
		\caption[Virtual sensing time series prediction]{The prediction quality of the presented models is visualized using short time series examples of a service load from the test dataset. Very large amplitudes (top) benefit the most from LSTM network predictions, since the system stiffness is highly non-linear in regions of high absolute force. Average (middle) amplitudes are generally predicted with a very high accuracy by all models. Small signal offsets in the FRF model are especially noticeable for small oscillations (bottom).}
		\label{fig:VS_time_series}
	\end{center}
\end{figure}

Apart from its application in forward prediction, the proposed hybrid methods can also be applied to virtual sensing. Here, the three displacement sensor measurements are used to predict the three channels of force data. From a physical point of view, this task is different from FP, since it only relates system output quantities to each other as depicted in \autoref{fig:VSFP} of \autoref{sec:experimental_setup}, while ignoring the drive signal. From a data-driven perspective, however, this virtual sensing example simply requires different measurements from the dataset, while the algorithms remain unchanged.\\
The general procedure of hyper-parameter identification is performed as described in \autoref{sec:forward_prediction_evaluation}. Like during FP, the best overall results were achieved by using a subsequence length $L$ of 256 and an overlap factor $o$ of 0.5. The chosen pure LSTM model features a single memory block with 29 memory cells and was trained for 501 epochs using a learning rate $\lambda$ of 0.0002. Both hybrid models use one memory block with 39 cells, where \textit{hybrid~1} was trained for 402 epochs with $\lambda=0.0001$ and \textit{hybrid~2} was trained for 501 epochs with $\lambda=0.0002$.\\
The models are compared by evaluating a variety of metrics using the noise and service load files of the test dataset, visualized in \autoref{fig:VS_results}. Like in the forward prediction task, the test bench drive signals are rescaled from the same original signal shape using different scaling factors between 0.3 and 1, and the negative and positive offsets lead to increasing non-linearity of the system behavior.
\\
Regarding the noise files of the testing dataset, the observations are very similar to the FP results. The mixed noise signal N1 is predicted with very high accuracy by all LSTM based approaches, showcasing the ability of these models to interpolate in the training dataset. On the other hand, the noise signal N2 is composed of data points at the boundary of the LSTM model input space, which are not included in the training dataset. Here, the predictions of LSTM based models are very poor. In the service load examples, the linear FRF model is outpreformed significantly by both pure LSTM and hybrid approaches. The choice of the best model depends on which metric is most important for the application case. For fatigue life estimations, the Multi-Rain damage ratio is very important, which is best approximated by the \textit{hybrid~2} approach. However, if the approximation quality for the power spectral densities is of primary interest, pure LSTM network and \textit{hybrid~1} should be selected.\\
Exemplarily, \autoref{fig:VS_time_series} provides a visual time series comparison of the model predictions for a single force channel.

\subsection{Dataset size dependency}
\begin{figure}
	\begin{center}
		\subfloat{
			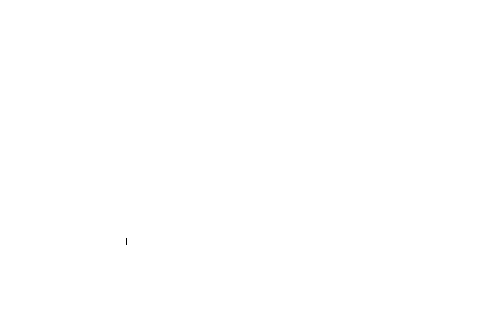
		}\\
		\subfloat{
			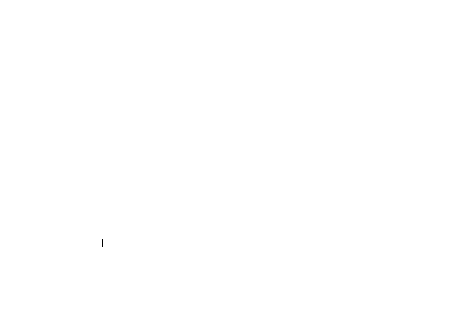
		}
		\caption[Linear study]{The influence of the dataset size on the predictive quality of LSTM networks is studied using a large dataset of random noise excitations and synthetic FRF model responses. The comparison of different model architectures shows that very small models are not able to properly exploit the benefits of a large dataset, while the prediction quality of more complex models scales very well with the dataset size.}
		\label{fig:linear_study_results}
	\end{center}
\end{figure}

The prediction quality of data-driven algorithms strongly depends on the availability of training data. Unfortunately, knowledge about this dependency is rarely available in practical applications, since the collection of large datasets can be very expensive and time consuming. In order to estimate how well LSTM network predictions scale to both larger and smaller datasets, an additional study was conducted using synthetic data.\\
In order to facilitate the process of training data generation, linear time invariant system dynamics are assumed for this study. The relation between input and output channels is therefore perfectly captured by an FRF model. This model can be used to generate arbitrary amounts of synthetic output data from randomly generated input data for the LSTM network parameterization.\\
In this study, the FRF model of the virtual sensing experiment in \autoref{sec:virtual_sensing_evaluation} is used. For training and validation purposes, 250 data files of uncorrelated white pink noise data are generated with a length of \SI{180}{\second} each. The white section is limited at \SI{20}{\hertz}, the following pink section is characterized by a power spectral density of the form
\begin{align}
	\text{PSD}(\omega) \propto \omega^{-1}
\end{align}
and stops at \SI{50}{\hertz}. The channel mean is randomly sampled in the range [\SI{-4}{\kilo \newton},\SI{4}{\kilo \newton}] and the amplitude range is [\SI{0.5}{\kilo \newton},\SI{2}{\kilo \newton}].\\
Four different LSTM architectures, shown in \autoref{tab:linear_study_models}, are compared in this study. Here, $\{A\}$ or $\{A, B\}$ denote the architecture of a network, where $A$ and $B$ are the number of memory cells in the first and second block, respectively.
\begin{table*}
	\centering
	\caption[Linear study networks]{Network architectures in the linear study}
	\label{tab:linear_study_models}
	\begin{tabular}{ p{3.5cm} P{1.5cm} P{1.5cm} P{1.5cm} P{1.5cm} }
		\toprule
		Architecture & \{10\} & \{39\} & \{23,23\} & \{39,39\} \\
		\midrule
		Number of parameters & 593 & 6,828 & 6,880 & 19,152 \\
		\bottomrule
	\end{tabular}
\end{table*}
The network with one block of only 10 inner states is chosen as an example with an exceptionally low degree of freedom. The architecture of the following networks with one block of 39 cells or two blocks of 23 cells are each very similar to the best performing networks of the forward prediction and virtual sensing tasks, see \autoref{sec:forward_prediction_evaluation} and \autoref{sec:virtual_sensing_evaluation}, respectively. The final network with two blocks of 39 memory cells each is considerably more complex than any previously examined architecture.\\
Since this study is designed to investigate the influence of the dataset size on the LSTM prediction, the model hyper-parameters are fixed to a subsequence length $L$ of 256, an overlap factor $o$ of 0.5, a learning rate $\lambda$ of 0.001 and a fixed training length of 100 epochs. The number of randomly generated noise training files was varied between 5 and 200, where each file has a length of \SI{180}{\second} and was sampled at \SI{1}{\kilo\hertz}. For each model architecture, one random initialization was generated and used as the starting point for all dataset sizes. The trained models are compared on a dataset of 50 random noise signals, which are independent from the training data. \autoref{fig:linear_study_results} shows the RMS and Multi-Rain damage ratio error metrics, averaged over this validation dataset.
\\
The RMS error of each model improves with an increasing number of training data files. Similarly, the Multi-Rain damage ratio converges towards the perfect result of 1. It is also apparent that the small model with only 10 memory cells is not able to capture the system characteristics as good as larger models. Although the model with two blocks of 39 cells performs best, it does not yield significantly better predictions than the models with architectures [39] and [23,23] while requiring substantially more computational effort to parameterize.

\section{Discussion}
In \autoref{sec:forward_prediction_evaluation} and \autoref{sec:virtual_sensing_evaluation}, it was shown that hybrid modeling strategies can greatly improve upon the prediction accuracy of FRF models in system identification tasks for FP and VS. Interestingly, the pure LSTM network did not outperform the FRF model in the FP task, despite being inherently non-linear and using many more training data samples, which cover the system behavior as a whole much better. A possible explanation for the improved performance of hybrid models compared to pure LSTM networks is that the FRF predictions reduce the model complexity required from the LSTM network, since important linear system characteristics are directly available. Because high model complexity generally requires an increased amount of training data, it can be concluded that hybrid models use the available data more efficiently than pure neural network based models. In direct comparison, the \textit{hybrid~2} approach of extending the LSTM network inputs by the FRF prediction generally yields more accurate and reliable results than the \textit{hybrid~1} approach of using the LSTM network to correct the FRF model error. This was especially notable regarding the Multi-Rain damage ratio fatigue metric.\\
The evaluation results of noise and service load testing data also showed that both signal shapes achieve an overall comparable accuracy. This is important, since it means that the general approach of using noise data for parameterization and deploying the models for service load data predictions is feasible. From the poor prediction results of noise file N2, which featured input data ranges that were not present during training, it can be assumed that good coverage of the boundary regions of the system behavior is very important for neural network based models.\\
In general, it must be emphasized that all LSTM models are trained individually from different initializations. This has implications, which can be seen in the RMS error metric of \autoref{fig:FP_results} for the displacement prediction of N2. Because of the individual initialization and training history, it is possible that a hybrid model performs worse than pure LSTM and FRF models in some specific regions of the system behavior, even though hybrid models perform far better in most situations.\\

\section{Conclusions} \label{sec:conclusions}
A novel hybrid approach for forward prediction and virtual sensing in non-linear dynamic systems was introduced based on Long Short-Term Memory networks and frequency response function models. These methods synergize very well, since the FRF model perfectly captures the behavior of linear systems and therefore provides a very good starting point for the non-linear LSTM predictions. Extraction and recombination of short subsequence signals enable the direct application of the LSTM algorithm to measurement data. Two strategies for the hybrid combination of LSTM and FRF models were suggested and compared. The effectiveness of the proposed approaches was demonstrated using a non-linear experimental dataset from a servo-hydraulic fatigue test bench. Different error metrics were employed to determine the predictive quality of the models in time and frequency domains as well as in the context of fatigue life assessment. Linear studies on a large synthetic dataset suggest that further improvements in prediction quality are likely to be achieved by increasing the dataset size.\\
The presented forward prediction models can be incorporated into existing workflows for dynamic response simulation. By simulating fatigue test rig responses, both the pre-damage in test specimen and the total energy consumption of fatigue tests can be reduced. While LSTM models are parameterized efficiently from noise data and provide very fast predictions, high computational efforts are required to conduct hyperparameter studies. Fortunately, the network training process profits immensely from parallelization, which reduces this issue. Still, strategies to accelerate the training process should be investigated in further research. Additionally, it would be very beneficial to identify algorithms which find an optimal dataset composition for training and validation purposes. Such algorithms could estimate the amount of data required to parameterize models for specific tasks, ensuring that the available measurements are used as efficiently as possible.\\
This paper motivated the potential of hybrid models for virtual sensing application using system responses of a fatigue test bench. For structural health monitoring or predictive maintenance applications in real systems, the sensor setup has to be adapted to the fatigue related physical quantities of interest. Especially low cost acceleration sensors are readily available and can provide important information on the dynamic state of vehicles, structures or industrial machinery in general. It should also be investigated how well the LSTM training process scales to scenarios with a significantly higher number of sensors.

\section{Declaration of Competing Interrest}
The authors declare that they have no known competing financial interests or personal relationships that could have appeared to influence the work reported in this paper.

\section{Acknowledgments}
\begin{table}[H]
 \begin{tabular}{m{8cm} m{1.5cm} m{5cm}  }
  \includegraphics[height=20mm]{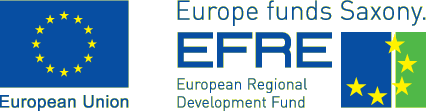} &   \includegraphics[height=20mm]{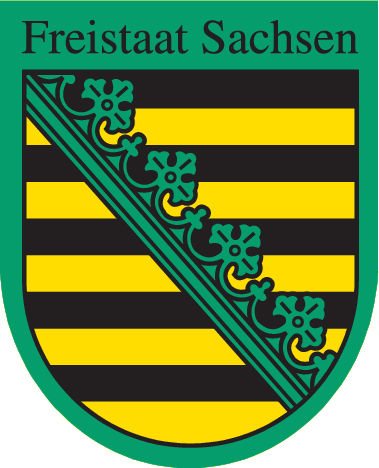} & 
This measure is co-financed with tax revenues on the basis of the budget adopted by the members of the Saxon State Parliament.
 \end{tabular}
\end{table}	
The authors gratefully acknowledge the GWK support for funding this project by providing computing time through the Center for Information Services and HPC (ZIH) at TU Dresden.

\bibliography{Bibliography}

\end{document}